\documentclass[prl,amsfonts,twocolumn]{revtex4}
\usepackage{amsmath,bm}    
\usepackage{amssymb}
\usepackage{amsthm}
\usepackage{wasysym}

\usepackage{bm}

\usepackage{amscd}
\usepackage{graphicx}   
\usepackage{verbatim}   
\usepackage{color}      
\usepackage{subfigure}  
\usepackage{hyperref}   
\usepackage{dsfont}

\usepackage{mathtools}

\newcommand{\Trace}{\operatorname{Tr}}

\DeclareMathOperator*{\Motimes}{\text{\raisebox{0.25ex}{\scalebox{0.8}{$\bigotimes$}}}}

\def \HilbL{\mathcal{H}}
\def \q{\mathbf{q}}

\def \Ue{\UI^{e}}

\def \Ucirc{U_{\mathrm{circ}}}
\def \Ugate{U_{\mathrm{gate}}}

\def \T{\mathbf{T}}

\def \HK{H_{\scriptscriptstyle  K}}
\def \HI{H_{\scriptscriptstyle  I}}

\def \UK{U_{\scriptscriptstyle  K}}
\def \UI{U_{\scriptscriptstyle I}}

\def \TU{\tilde{U}}
  
 \def   \LNT{\mathcal{L}_{\scriptscriptstyle  N \times 2t}}

\begin{document}
\title{ Exact  local correlations  in     kicked chains at light cone edges}

\author{Boris Gutkin$^2$, Petr Braun$^1$, Maram Akila$^{1,3}$, Daniel Waltner$^1$, Thomas Guhr$^1$}
\affiliation{$^1$Fakult\"at f\"ur Physik, Universit\"at Duisburg-Essen,
  Lotharstra\ss e 1, 47048 Duisburg, Germany\\
  $^2$Department of Applied Mathematics, Holon Institute of Technology, 58102 Holon,
Israel\\
$^3$Fraunhofer IAIS, Schloss Birlinghoven, 53757 Sankt Augustin, Germany}

\begin{abstract}  

We show that local correlators   in   a wide class of  kicked chains can be calculated exactly  at light cone edges.   Extending previous works on dual-unitary systems, the   correlators between local operators    are expressed through the expectation values of   transfer matrices  $\T$ with  small  dimensions. Contrary to the previous studies, our results are not restricted to dual-unitary systems with  spatial-temporal symmetry of the dynamics. They   hold for a generic case without fine tuning  of  model    parameters.  The results     are  exemplified on the  kicked Ising spin chain model, where we provide an explicit formula for  two-point correlators near light cone edges beyond the dual-unitary regime.

 
\end{abstract}

\maketitle

\textit{Introduction.}  Spatially extended Hamiltonian systems with
 local interactions are paradigm systems in the field of
 many-body physics. On the experimental side, various aspects become
 ever better amenable to direct measurement \cite{BDZ08, Mahan11,
 Schreiber842, Simon2011} whilst a recent burst of activities
 \cite{Engl2014, Dubertrand_2016, Abanin2015,Atas_2014,Keating2015,
 Czischek_2018} greatly improved our theoretical understanding.  In
 the context to be addressed here, the outstanding importance of these
 systems is rooted in their spatiotemporal correlation of local
 observables which describe, in an often generic manner,
 experimentally accessible features of interacting many-body systems
 such as spectral statistics or transport properties \cite{AS10,
 Sethna06, Mahan11}. 
 The wealth of available results, unfortunately, covers systems which
 are either  dynamically too simple, such as free or integrable ones, or too low in
 dimension, such as  cat or baker maps.  It is thus of paramount interest to find
 representatives of those systems capturing, on the one hand, the full
 complexity and, on the other hand, allowing for analytical treatment.


 In this work we consider a class of systems admitting   a number of
 different dynamical descriptions \cite{AWGG16}. The standard one
 corresponds to the system evolution with respect to time, induced by
 the system Hamiltonian.  Alternatively, one can consider evolution
 along one of the spatial directions. In this  \textit{dual}
 approach the
 corresponding coordinate takes on the role of time. The resulting
 dynamical system is generically a non-Hamiltonian one \cite{AWGG16,
 AWGBG16, AGBWG18}.  However, in some special cases it might happen
 that the dual spatial evolution is a Hamiltonian one, as well.  The
 representatives of such systems, referred to as
 \textit{dual-unitary}, can be found among coupled map lattices
 \cite{GutOsi15, GHJSC16}, kicked spin chains \cite{BeKoPr18,
 BeKoPr19-1, BWAGG19, LakshPal2018}, circuit lattices
 \cite{BeKoPr19-4, GopLam19, BeKoPr2019operator} and continuous field
 theories \cite{AVAN2016}.

   Dual-unitary systems   have recently attracted   considerable attention \cite{BeKoPr19-4, GopLam19, BeKoPr19-1, LakshPal2018, BeKoPr18, BWAGG19,BeKoPrPi19, BeKoPr2019operator, KrPr2019KPZ,zhou2019entanglement,AVAN2016,Karl15,Arul19} due to their      intriguing properties.
  On the one hand,   these models generically exhibit features of  maximally chaotic many-body systems. In particular, their spectral statistics are   well described by the Wigner-Dyson distribution.  They  are  insusceptible to  many-body localisation  effects   even in the presence of strong disorder \cite{BeKoPr18, BWAGG19}. The  entanglement has been shown to grow linearly  with time and to  saturate   the maximum bound. 
  On the other hand,  dual-unitary  models turned out to be amenable to exact analytical treatment. The growth of the entanglement  entropy for kicked Ising spin chains (KIC) for certain types of initial states has been evaluated exactly in \cite{BeKoPr19-1} and their entanglement spectrum was found  to be trivial \cite{GopLam19}.

  It has been recently shown  that two-point  correlations of   local operators in  dual-unitary  quantum circuit   latices  \cite{BeKoPr19-4,  CL20} and kicked chains (KC) \cite{GBAWG20} can be expressed exactly   in terms of small dimensional transfer operators. The main goal of the present contribution is to demonstrate that, in fact, an analogous result holds in a much more general setting. We consider here    KC  built upon a pair of $L\times L$   matrices $u_1, u_2$. The  model  is  defined for an  arbitrary length $N$ of the chain  and an  on-site Hilbert  space dimension $L$. 
  It  becomes a dual-unitary one  when $u_1, u_2$ are complex Hadamard matrices with all entries having the same absolute values.
  
  In the body of the paper we show that
  correlators  between    local   operators along the light-cone edges can be expressed through the expectation values  of a  transfer matrix  $\T$  whose  dimension is determined by $L$ rather than $N$. This result does not rely upon  dual-unitarity and holds for generic model parameters. For  the dual-unitary case the correlators, furthermore, vanish outside of the light-cone edges, in agreement with  \cite{BeKoPr19-4, GBAWG20}. We illustrate our results on the example of KIC, where    we provide  an  explicit  formula  for  two-point local  correlators  at  light-cone  edges   outside of the dual-unitary regime.

\textit{Kicked chains (KC).} 
 In this paper we consider cyclic  chains of $N$
 locally interacting particles,  periodically kicked  with an on-site external potential.  The system is governed by the  Hamiltonian,
\begin{equation}
H(t)= \HI+\HK\sum_{m=-\infty}^{+\infty}\delta(t-m),\label{KickedChain}
\end{equation}
with $\HI$, $\HK$  being the  interaction and kick parts, respectively.   
The corresponding  Floquet time evolution is the product  of the operators,  $\UI=e^{-i\HI}$ and   $\UK= e^{-i \HK}$,   acting on the Hilbert space  $\HilbL^{\otimes N}$  of  the dimension $L^N$, where    $\HilbL = \mathbb{  C}^L$ is  the local Hilbert space equipped   with the basis $\{|s\rangle, s=1,\dots, L\} $. We require  that   $\HI$ couples nearest-neighbour sites of the chain  taking on a   diagonal form   in the product basis,  $\{|\bm s\rangle=|s_1\rangle|s_2\rangle \dots|s_N\rangle\}$.  The respective evolution is   fixed by  a   real function $f_1$, 
\begin{equation}
\langle \bm s|\UI[f_1]|\bm{ s'}\rangle=\delta(\bm s ,\bm{s'}) e^{i \sum_{n=1}^N f_1(s_n,s_{n+1})},
\end{equation}
with $\delta(\bm s ,\bm{s'}) =\prod_{i=1}^N \delta(s_i-s_i')$, 
and  cyclic boundary condition $s_{N+1}\equiv s_1$.
The second, kick  part, is given by the tensor product
\begin{equation}
\UK[f_2]=\Motimes_{i=1}^N u_2,  \langle \bm s|\UK[f_2]|\bm{ s'}\rangle=\prod_{i=1}^N \langle s_i|u_2|{ s}'_i\rangle,
\end{equation}
where  $u_2$ is a  $L\times L$  unitary  matrix with the    elements 
$e^{i f_2(n,m)}/\sqrt{L}$ determined by a complex  function $f_2$.
Combining the two parts together we obtain the quantum evolution 
\begin{equation}
U= \UI[f_1] \UK[f_2],
\label{eq:basePropagator}
\end{equation}
acting on the Hilbert space of dimension $L^N$.

In the same way one constructs the  dual evolution acting  on the Hilbert space of dimension $L^T$ by exchanging $N\leftrightarrow T$ and $f_1 \leftrightarrow f_2$:  
 \begin{equation}
 \TU= \UI[f_2] \UK[f_1].
\end{equation}
 The following remarkable duality relation  \cite{AWGBG16,AGBWG18} holds between their traces for any integers $T$, $N$:
\begin{equation}
\mbox{Tr } U^{T}=\mbox{Tr }  \tilde{U}^{N}.
\end{equation} In contrast to the original evolution,  $\TU$ is a  non-unitary operator, in general. However, if   
\begin{equation}
\langle n|u_1|m\rangle=\frac{e^{i f_1(n,m)}}{\sqrt{L}},   \langle n|u_2|m\rangle=\frac{e^{i f_2(n,m)}}{\sqrt{L}},
\end{equation}
are  $L\times L$ complex Hadamard matrices (i.e., unitary matrices which matrix elements have the same  absolute value)  the dual operator, $\TU$ is unitary as well.   
We refer to such models as dual-unitary. Note that in the dual-unitary case  both $f_1, f_2$ are real.    A wide  family  of   such models, referred as dual-unitary Fourier transform  chains  (FTC) were  constructed in     \cite{GBAWG20} for each $L$ by fixing  $u_1, u_2$ to be  the unitary discrete Fourier transform    multiplied   on both sides by arbitrary diagonal unitary  and permutation matrices.
The correlators in  dual-unitary KC  have been  studied in \cite{GBAWG20}. Here we are primarily  focused   on  general  KC models with no  demand of dual-unitarity.  
\

 \textit{Correlations between  local   operators.}
 Let $(\q_1, \q_2 )$,  $(\q_3, \q_4 )$ be   two  pairs   of   matrices acting on the on-site Hilbert space $ \HilbL$. We define the corresponding many-body operators
  \begin{equation}\Sigma_{n_1}=\underbrace{\mathds{1}\otimes  \dots\otimes \mathds{1}}_{n_1 -1} \otimes\, \q_1\otimes\q_2\otimes \underbrace{\mathds{1}\otimes  \dots\otimes \mathds{1}}_{N-n_1-1 }\label{operator2}\end{equation}
 
 \begin{equation}\Sigma_{n_2}=\underbrace{\mathds{1}\otimes  \dots\otimes \mathds{1}}_{n_2 -1} \otimes\, \q_3\otimes\q_4\otimes \underbrace{\mathds{1}\otimes  \dots\otimes \mathds{1}}_{N-n_2-1 }\label{operator1}\end{equation}
  supported at the sites  $n_1,n_1+1$ and  $n_2, n_2+1$   of the chain, respectively.
  In what follows we  consider  the two-point correlator:
\begin{equation}
    C(n,t)=L^{-N}\Trace U^{t} \Sigma_{n_1} U^{-t} \Sigma_{n_2},\label{eq:fourpointCorr}
\end{equation}
where we assume $n=n_2-n_1 >0, t>0$. By  translation symmetry of the model,  we can set  $n_1=1$, $n_2=n+1$ without  loss of generality.

The above correlation function  can be written in the form of the partition function, 
\begin{multline} 
C(n,t)=\frac{1}{L^{Nt}}
    \sum_{\{s_{mk}| (m,k)\in\mathcal{L}_1\}}e^{-i\mathcal{F}(\{s_{m k}\})}  \\
   \left[ \prod_{(m,k)\in\mathcal{L}_2} \!\!\!\!\!\!\delta(s_{mk},s_{m,2t-k+1})\right]  D(s_{n_1 1},\dots  ,s_{n_2 t}),\label{method1body}
\end{multline}
where the last factor, \begin{multline}D= \langle s_{n_1,2t}| \q_1^c |s_{n_1,1}\rangle \langle s_{n_1,2t}| \q_2^c |s_{n_1,1}\rangle\\
\langle s_{n_2,t}| \q_3 |s_{n_2,t+1}\rangle\langle s_{n_2,t}| \q_4 |s_{n_2,t+1}\rangle, \end{multline}  $\q_1^c=u_2 {\q}_1 u^\dagger_2$, $\q_2^c=u_2 {\q}_2 u^\dagger_2$, depends on the eight  lattice sites, $\mathcal{L}_0=\{(n_1,k), (n_1+1,k)| k = 1, 2t\}\cup \{(n_2,k), (n_2+1,k)| k = t, t+1\}$  corresponding to the location  of the  observables and  the function $\mathcal{F}(\{s_{m,k}\})$ is given by eq.~(\ref{part_func2}) in the supplementary material.  The above sum  runs over  $2t\times N$ sites  of the  lattice $\mathcal{L}_1=\{(m,k)| k=1, \dots ,2t, m=1,\dots,N\}$ while  the product in (\ref{method1body}) is, furthermore,  restricted to the  subset   $\mathcal{L}_2=\{(m,k)| k=1,t, t+1, 2t, m=1,\dots,N\} \setminus \mathcal{L}_0$.   

\begin{figure}
    \centering
    \includegraphics[width=0.455\textwidth]{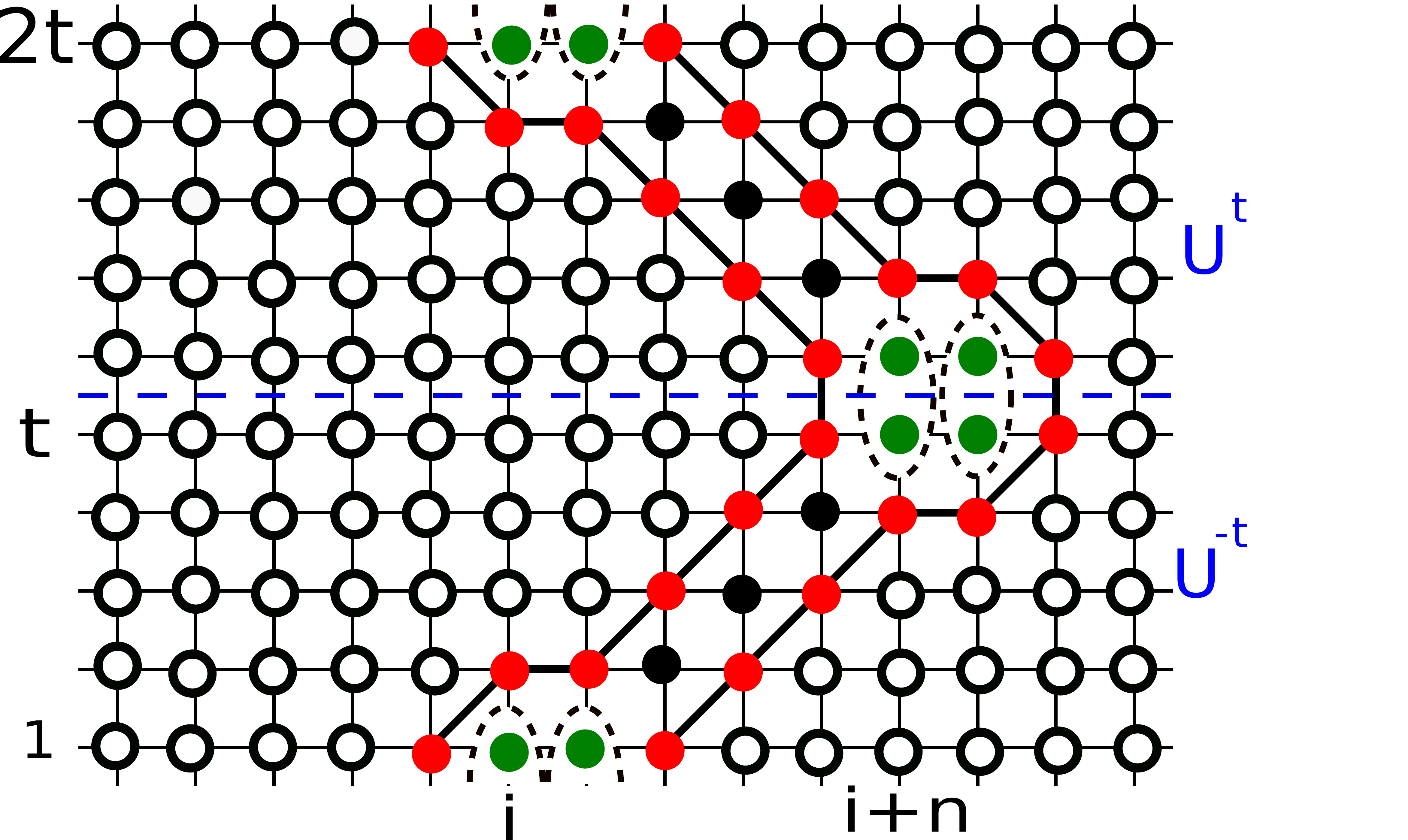}
    \caption{Elimination  of the summation variables in the partition function (\ref{method1body})  representing 4-point correlator (\ref{eq:fourpointCorr}).  The eliminated sites are shown   by empty black  circles. The remaining sum along the light-cone edge can be represented  in the form of the  expectation value (\ref{four_point_corr5})
    of the  $L^2\times L^2$  transfer  operator $\T$. }
    \label{fig:four_point2}
\end{figure}

The partition function (\ref{method1body}) allows for an instructive   graphical representation  illustrated on fig.~\ref{fig:four_point1}. 
As we show in the supplementary  material,  on the light cone edge, $n=t$  it  can be considerably simplified by eliminating   most of the variables, $s_{n,t}$ provided that  $N>2t$. The remaining lattice sum contains only the variables along the light cone edges as shown on fig~\ref{fig:four_point2}.  The resulting    expression  
can be represented in the form of  the  expectation value
\begin{equation}
    C_t\equiv C(t,t)=\langle\bar{\Phi}_{\q_1\q_2}|\T^{t-2}|\Phi_{\q_3\q_4}\rangle,
\label{four_point_corr5}
\end{equation}
of 
the transfer operator $\T$,
\begin{multline}
   \langle\nu \eta|\T|\eta'\nu'\rangle=\\
   \frac{1}{L^3}\left|\sum_{s=1}^L e^{i(f_1(\eta,s)+f_1(s,\nu') +f_2(\nu,s)+f_2(s,\eta'))}\right|^2 ,\label{transferOp}
\end{multline}
  acting on  the small  space $\HilbL\otimes \HilbL$.
The left $\bar{\Phi}_{\q_1\q_2}$ and the right  $\Phi_{\q_3\q_4}$  vectors are defined as 
 \begin{equation}
   \langle\nu \eta|\Phi_{\q_3\q_4}\rangle=
\frac{1}{L^3}\sum_{a,\bar{a},b=1}^L  \Gamma_{a\bar{a}}^b\langle a|\q_3|\bar{a} \rangle \langle b|\q_4|b\rangle,\label{vectors1}  
 \end{equation}
 \begin{equation}
\langle\bar{\Phi}_{{\q}_1{\q}_2}|\eta\nu\rangle=
\frac{1}{L^3}\sum_{a,\bar{a},b=1}^L
 \bar{\Gamma}_{a\bar{a}}^b\langle a| {\q}^c_2|\bar{a} \rangle \langle b|{\q}^c_1 |b\rangle, \label{vectors2}
\end{equation}
 where
 \begin{gather*}
 \Gamma_{a\bar{a}}^b=e^{i(f_1(\eta,\bar{a})-f_1(\eta, a) +f_2(\bar{a},\nu)-f^*_2(a,\nu)-f_1(a,b)+f_1(\bar{a},b))}  \\
\bar{\Gamma}_{a\bar{a}}^b=e^{i(f_1(a,\nu)-f_1(\bar{a},\nu) +f_2(\eta,a)-f_2^*(\eta,\bar{a})+f_1(b,a)-f_1(b,\bar{a}))}.
\end{gather*}

\begin{figure}
    \centering
    \includegraphics[width=0.455\textwidth]{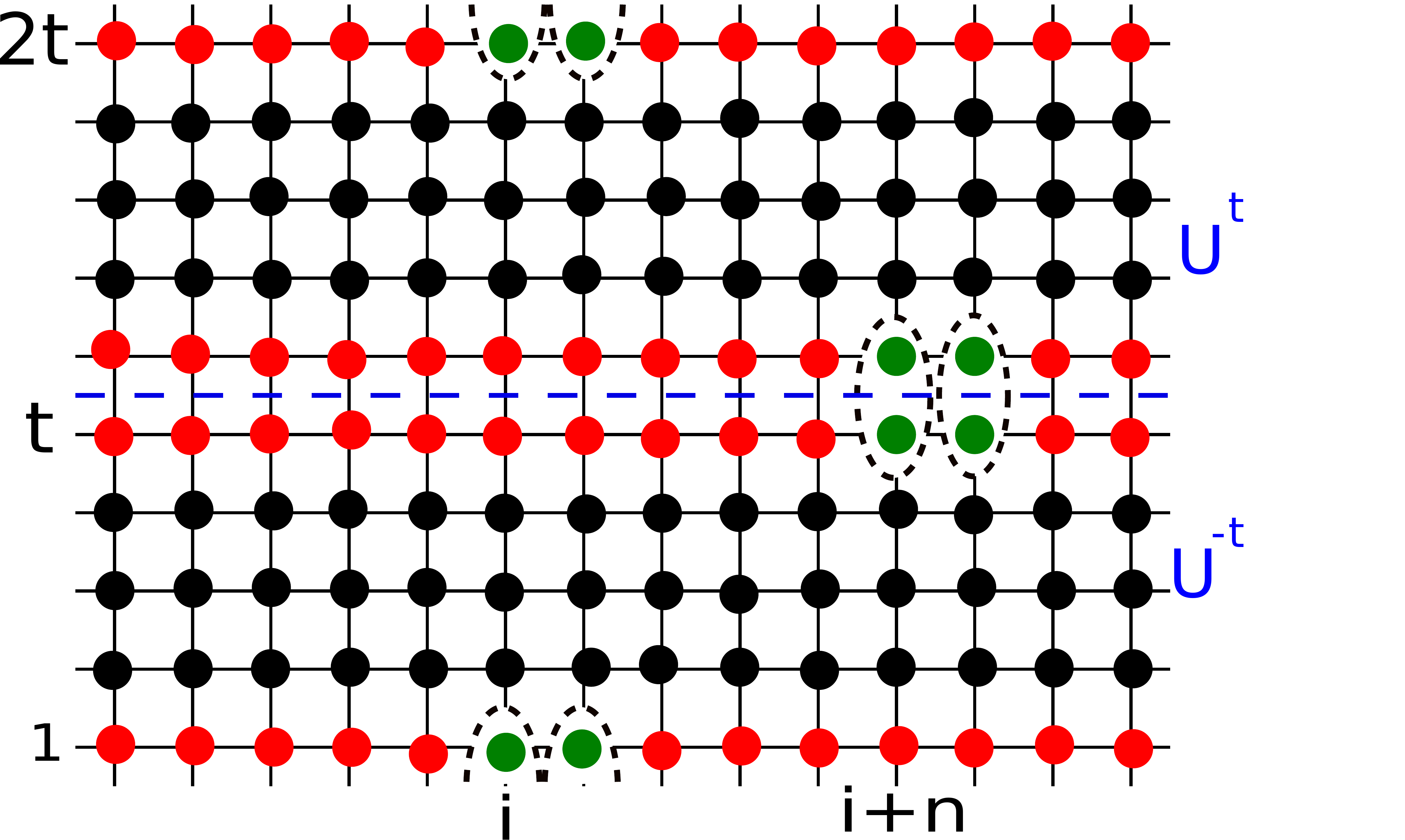}
    \caption{ The  picture  illustrates   the initial expression (\ref{method1body}), where the sum  runs  over $N\cdot 2t$ variables $s_{m,k}$. Circles in red  show $(m,k)$ sites, where    variables are    paired by the condition  $s_{m,k}=s_{m,2t-k+1}$. The green circles correspond to the location of the operators $\Sigma_{i}, U^{-t}\Sigma_{n+i} U^t$. }
    \label{fig:four_point1}
\end{figure}

It is easy to check that $\T$ is doubly stochastic  i.e, satisfies
\begin{equation*}
 \sum_{\nu=1}^L\sum_{\eta=1}^L\langle\nu \eta|\T|\eta'\nu'\rangle=\sum_{\nu'=1}^L\sum_{\eta'=1}^L\langle\nu \eta|\T|\eta'\nu'\rangle=1.
\end{equation*}
This implies  that the  spectrum of $\T$ is contained within the unit disc with   the largest  eigenvalue, $\mu_1=1$.  The left (resp. right) eigenvector corresponding to $\mu_1$ are  given  by the choice $\q_3=\q_{4} =\mathds{1}$ (resp. $\q_1=  \q_{2} =\mathds{1}$). 
For typical system parameters  the correlators between traceless observables decay exponentially with the   rates determined  by the second  eigenvalue $\mu_2$, $|\mu_2|\leq |\mu_1|$    of $\T$  having  the largest absolute value after $\mu_1$.  
Eq.~(\ref{four_point_corr5}) can be also used to evaluate correlations  between strictly local observables in KC by setting   $\q_1=\mathds{1},  \q_{4} =\mathds{1}$. While   $\bar{\Phi}_{\mathds{1}\q_2} = \Phi_{\q_3\mathds{1}}=0$ for traceless $\q_2,\q_3$ in dual-unitary KC, these vectors  do not vanish  generically, implying non-trivial correlations $C_t$ between  strictly local operators in a general  KC.

It is important to emphasize that (\ref{four_point_corr5}) holds for any KC   model (\ref{KickedChain}) and does not require dual-unitarity. In essence,  any KC of this type is solvable, as far as, local correlators are restricted to the light cone edge. What makes dual-unitary case  special is that  $C(n,t)$ is zero there for traceless $\q_i$'s if $n\neq t$ and  $N>2t$. 
 As has been pointed out in \cite{BeKoPr19-4}, this  can be understood in a simple intuitive way.    Since the   speed of information  propagation in KC   (\ref{KickedChain})  equals one,  the  correlator  of   operators (\ref{operator2},\ref{operator1})  with traceless $\q_i$'s     must vanish  outside of the light cone $ |t| < |n|$, $n=n_2-n_1$.  By   the dual unitarity, a similar result holds for  points within the light cone  $ |t| > |n|$, as well. This leaves the light cone edges $|t|=|n|$ as  the    only possible places on the space-time lattice  where non-trivial correlations might arise. Accordingly, for dual-unitary models we have 
 \begin{equation}
     C(n,t)=\delta(n,t) C_t, \label{another formula}
 \end{equation}
   where $C_t$ is given by eq.~(\ref{four_point_corr5}).

The  above results  can be straightforwardly extended to   systems with spatial-temporal disorder, where the local functions $f_1,f_2$ depend on the lattice sites. In such a case the transfer operator $\T^{t-2}$ in  eq.~(\ref{four_point_corr5}) is  substituted with  the product of  local ``gate" operators $\T_1\T_2\dots \T_{t-2}$, where each $\T_i$   is determined by the functions  $f_1, f_2$ at the  point  $(i,i)$ of the  spatial-temporal lattice. 
 For a sub-family  of dual-unitary, FTC models introduced in \cite{GBAWG20}  all matrices $\T_i$  are diagonalized by one and the same unitary transformation. As a result,  the decay exponents of the  correlators  (\ref{eq:fourpointCorr}) in the disordered FTC  are just given by the averages of  the local exponents.\

 \textit{ KIC model.} 
 Below we illustrate our results on the example of KIC model providing   a minimal,  $L=2$,  realisation of 
the KC   model (\ref{KickedChain}). The  KIC evolution is governed by the  Hamiltonians:
\begin{equation}
 \HI=\sum_{n=1}^N J \hat\sigma_n^z  \hat\sigma_{n+1}^z +h \hat\sigma_n^z,\quad 
\HK=b\sum_{n=1}^N   \hat\sigma_n^x,   \label{KICHamiltonian} 
\end{equation}
\[ \hat\sigma_n^\alpha=\underbrace{\mathds{1}\otimes  \dots\otimes \mathds{1}}_{n -1} \otimes\,  \sigma^\alpha\otimes \underbrace{\mathds{1}\otimes  \dots\otimes \mathds{1}}_{N-n },\]
where $\hat\sigma_1^\alpha=\hat\sigma_{N+1}^\alpha$ and  $\sigma^\alpha, \alpha=x,y,z$ are Pauli matrices.

For the sake of simplicity of exposition we set $b=\pi/4$ with $h$ and $J$  being arbitrary.  For this choice of parameters 
  eq. (\ref{four_point_corr5})   gives  (see supplementary material) for  the correlator (\ref{eq:fourpointCorr}) at $n=t$, $N>2t$: 
\begin{equation}
C_t=\mathcal{C}_{\alpha\beta}^{\gamma\delta}  (\sin^2 2J\cos 2h)^{t}, \label{main_result}
 \end{equation}
 where    
 the prefactors $\mathcal{C}_{\alpha\beta}^{\gamma\delta} $ depend  on the operators  $\q_1=\sigma^\alpha, \q_2=\sigma^\beta,\q_3=\sigma^\gamma, \q_4= \sigma^\delta$. Specifically,
 $
\mathcal{C}_{yz}^{yz}=1$, $ \mathcal{C}_{yx}^{xz} =\tan^2 2h,
$
 $\mathcal{C}_{yx}^{yz}= \mathcal{C}_{yz}^{xz} =-\tan 2h$
 and   zeroes for all other spin combinations.  The dual-unitary  case  corresponds to $J=b=\pi/4$ leading by (\ref{another formula}) to $C(n,t)=\delta(n,t)\mathcal{C}_{\alpha\beta}^{\gamma\delta}  (\cos 2h)^{t}$, the result obtained in \cite{GBAWG20}.

  As has been explained above, in the dual-unitary case  all  two-point corelators 
\begin{equation}
     C^{\alpha \beta}(n,t)=\frac{1}{2^N}\Trace \left(U^{-t} \hat\sigma_{n+1}^{\alpha} {U}^{t} \hat\sigma_1^{\beta}\right), \label{2_point}\end{equation}
 $\alpha,\beta\in\{x,y,z\}$ between local spin operators vanish identically for $t>0$, $N>2t$. 
For a general KIC, away from the self-dual regime, the correlators  (\ref{2_point})     are  non-zero, in general,  and  can be   evaluated at $n=t-1$, $N>2t$  by using 
  eq.~(\ref{four_point_corr5}). To this  end we set   $\q_1=\mathds{1},  \q_{4} =\mathds{1}$ and $\q_2=\sigma^{\alpha},  \q_{3} = \sigma^{\beta}$ which yields  for $t>1$
 \begin{equation}
  C^{\alpha \beta}(t-1,t)=\langle\bar{\Phi}_{\mathds{1}\q_2}|\T^{t-2}|\Phi_{\q_3\mathds{1}}\rangle.\label{four_point_corr2}\end{equation}
 For $b=\pi/4$ and general $J$ a straightforward evaluation of (\ref{four_point_corr2}) leads to
 \begin{equation}
 C^{\alpha \beta}(t-1,t)= \mathcal{C}^{\alpha \beta} (\cos 2h \sin^{2} 2J)^t\cot^2 2J \label{KICCorrelator}
 \end{equation}
 with the coefficients given by 
 \[
\mathcal{C}^{xx} = 1,\, \mathcal{C}^{xy} = \mathcal{C}^{zx} = \tan 2h,\, \mathcal{C}^{zy} = \tan^2 2h,
\]
and by  zeroes for other $\alpha,\beta$  pairs. 
Note that for all $n\geq t$ the correlator $C^{\alpha \beta}(n,t)$ vanishes. For $n=t$ this result  can be obtained by   the substitution   $\q_2=\mathds{1},  \q_{4} =\mathds{1}$ and $\q_1=\sigma^{\alpha},  \q_{3} = \sigma^{\beta}$ into (\ref{four_point_corr5}).  Since
$ \bar{\Phi}_{\q_1\mathds{1}}=0$, one gets immediately  $C^{\alpha \beta}(t,t)=0$. For a larger $n>t$, the same answer follows straightforwardly  from the fact that speed of information  propagation in KIC is one.
 
\begin{figure}
    \includegraphics[width=0.4\textwidth]{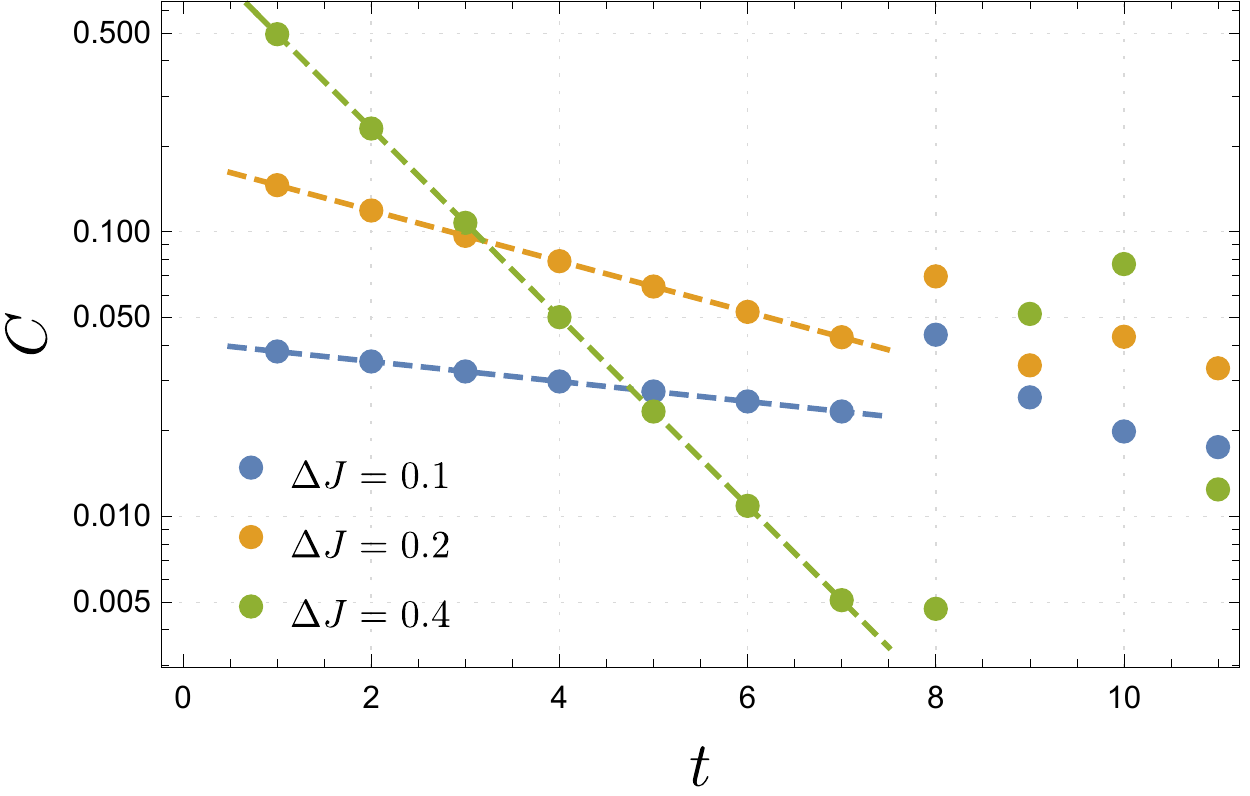}
    \caption[Decay of the correlator under perturbation of self-duality.]{
    Time dependence of $\frac{1}{2^N}\Trace \big(U^{-t} \sigma_t^x {U}^{t}$ $ \sigma_1^x\big) $ for  $N=14$ spins with  generic values of $J,h$ and $b=\pi/4$.  Straight lines are determined by eq.~(\ref{KICCorrelator}) with $h=3.0$ and  $ \Delta J=\pi/4-J=0.1,\, 0.2,\, 0.4$.
    The dots  are obtained by direct numerical calculation of  correlators for the same system parameters. Note perfect agreement with the analytic predictions for $
    2t\leq N$. For $2t >N$   eq.~(\ref{KICCorrelator}) is no longer valid which can be clearly observed at the plot.
    
    }
    \label{fig:PaperCorr}
\end{figure}

 The correlators (\ref{main_result}, \ref{KICCorrelator}) decay  exponentially with the rates $\cos2h \sin^2 2J$, except the cases  where $\frac{2J}{\pi}-\frac{1}{2}\in\mathbb{Z}$,  $\frac{2h}{\pi}\in\mathbb{Z}$, see fig.~\ref{fig:PaperCorr}.  For these parameters  KIC   corresponds to  well known cases of    integrable classical 2-d Ising spin model  with complex parameters \cite{LeeYangI,LeeYangII, Matveev_2008}.

 \textit{Conclusions.}
 We  derived   an analytic  formula, relating    correlators $C(n,t)$ between  operators with two point support  for  $n=t$ (light cone edge)   to the expectation values of a transfer operator with small dimensions.  The result holds for a sufficiently long  generic KC and does not require  fine-tuned  system parameters. 
 For the subfamily of  dual-unitary  KC this  allows for  a full characterization of the correlator behavior, as   $C(n,t)=0$  for $n\neq t$ in this case.    We illustrated  these results  on the example of KIC, where  an explicit expression for      correlations between strictly local operators has been obtained also next to  the light cone edge at $n=t-1$.  
 
 Our study  clarifies  the role of dual-unitarity with regard to the model  solvability. The fact, that  local correlators  in the vicinity of  the light cone edge can be expressed in terms of a small dimensional transfer operator  is due to the locality of the  system interactions. On its own it does not require  dual-unitarity of the system dynamics. The dual-unitarity is only essential to ensure that correlators of traceless operators vanish outside of the line $n=t$. For a general model one has $C(n,t)=0$  only  for $n>t$.       
 
 The above  results  allow for  several generalizations. First, models with a larger range of interactions can be treated in a similar manner. For systems with $r$-point interactions,
 $H_I=\sum_{i=1}^N f_1 (s_{1+i},\dots,s_{r+i})$  the correlations at  the light cone edge $n=rt$ can be expressed through  transfer operators of 
 the dimension $L^r\times L^r$. Second, in the present  work  we restricted  our considerations to  correlators  between operators with  two-point support. An analogous result holds  for  correlations between operators with a larger   support, i.e., $\Sigma^{(l)}_k=\mathds{1} \otimes \dots \otimes \mathds{1}\otimes\q_{k+1}\otimes\dots \otimes\q_{k+l}\otimes  \mathds{1} \otimes   \dots\otimes \mathds{1}$. 
 In general, the correlators $ \langle \Sigma^{(l)}_0 \Sigma^{(l)}_n\rangle$ can be expressed through    expectation values of  transfer operators $\T_l$ with the  dimensions $L^l\times L^l$.   By using this,  the correlators  $C(n,t)$ in (\ref{eq:fourpointCorr})   can be   evaluated   above  the light cone edge  $t=n+l$,   $l>0$  as well.  To this end one fixes all $\q_i$ in  $  \Sigma^{(l)}_0, \Sigma^{(l)}_n$ to $\mathds{1}$, except $\q_l, \q_{n+l}$.  The price to pay is  in the dimension of the   transfer operators -  the dimension of $\T_l$ increases exponentially with $l$. Finally, it is worth of noticing that for even $N$ and even  propagation times $t$  the  correlators  (\ref{eq:fourpointCorr})    can be mapped, in principle,  upon correlators of  a  circular lattice with a special gate operator $\Ugate$, provided by (\ref{gateOP})  in the supplementary material.    It  seems to be very plausible that an  analogue of our main result  (\ref{four_point_corr5}) holds for a general  circular lattice, as well.

\vspace{1em} 
\section*{Acknowledgements} We thank  T.~Prosen for  useful discussion.  One of us (B.G.) acknowledges  support from  the Israel Science Foundation through grant No.~2089/19.

\bibliographystyle{ieeetr}
\bibliography{references_kicLim_Corr}


\phantom{p. 1}
\clearpage

\onecolumngrid
\section{Supplementary material}
\subsection{Relation to  circuit lattices}

For the sake of  comparison   it is instructive to observe a  connection  between  quantum kicked chains  considered in this work  and  circuit lattices. Such a connection can be   established when both the  chain length $N$  and the propagation times $t$ are  even. It is straightforward to see that  the quantum evolution operator $U^{2t}$ for even times can be cast  into the form 
\begin{equation}
    U^{2t}=  \Ue \Ucirc^t (\Ue)^\dagger.
\end{equation}
Here, the operator $\Ue$ 
corresponds to the even  
``half of the  interaction":
\begin{equation}
\langle \bm s|\Ue[f_1]|\bm{ s'}\rangle=\delta(\bm s ,\bm{s'}) e^{i \sum_{n=1}^{N/2} f_1(s_{2n},s_{2n+1})},
\end{equation}
and the  evolution $\Ucirc$ has the form
\begin{equation}
\Ucirc= \mathbb{T} \Ue \UK \Ue \mathbb{T}^\dagger\Ue \UK\Ue,
\label{eq:circuitTrace}
\end{equation}
where $\mathbb{T}$ is the circular shift operator on a lattice of $N$ sites. Note that $\Ucirc$ has a special structure, 
characteristic to circuit lattice evolution, see e.g.,  \cite{BeKoPr19-4}. The role of the unitary gate operator is fulfilled here by 
\begin{equation}
\Ugate= u^e_1 \, (u_2\otimes u_2)\, u^e_1, \label{gateOP}
\end{equation}
where the diagonal matrix \[\langle  s_1 s_2|u^e_1|s'_1 s'_2 \rangle=\delta( s_1 ,{s'}_1)\delta( s_2 ,{s'}_2) e^{i  f_1(s_{1},s_{2})}\] 
is a restriction of $\Ue$ to two adjacent lattice sites.

By eq.~(\ref{eq:circuitTrace}) we find for the two-point correlator
\begin{equation}
\Trace\left( U^t Q_1 U^{-t} Q_2\right)=\Trace\left( \Ucirc^t \tilde{Q}_1 \Ucirc^{-t}\tilde{Q}_1 \right), 
\end{equation}
where $\tilde{Q}_i= (\Ue)^\dagger Q_i\Ue$. Since $\Ue$ couples two neighbouring sites,  any strictly local operator with one-point support  in the kicked model corresponds  to a two site operator of the respective  circuit model.

\subsection{ Graphical method for evaluation of correlators  }

 \begin{figure}
   \includegraphics[width=0.4\textwidth]{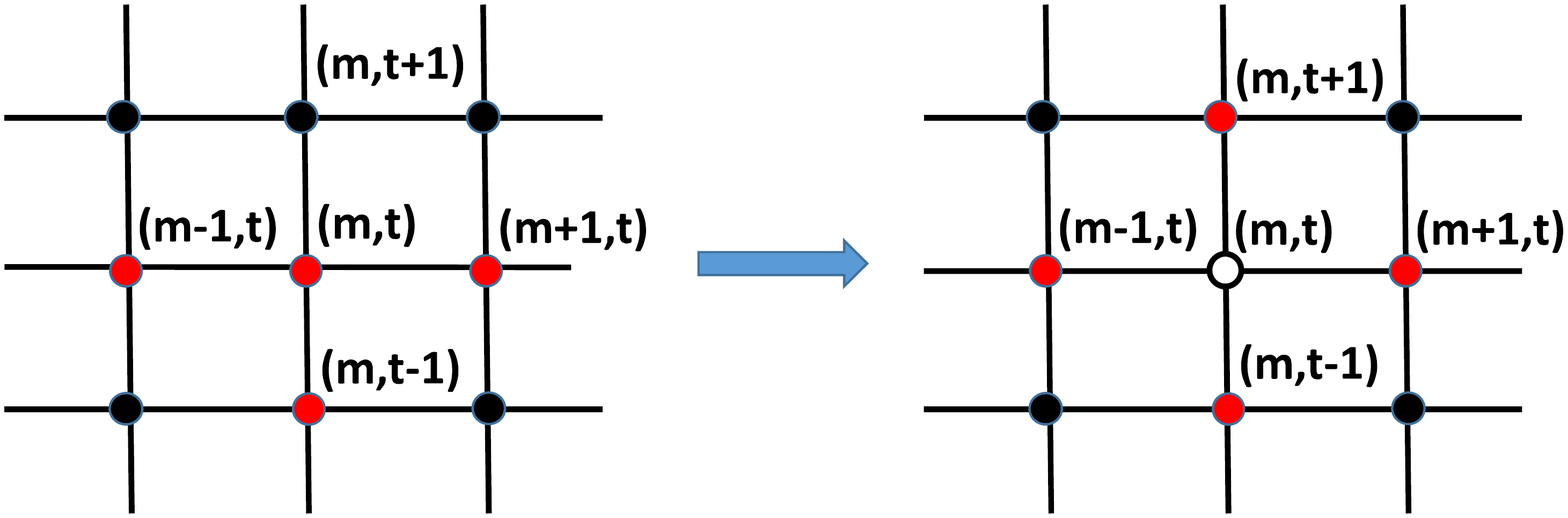}\hskip 2cm
   \includegraphics[width=0.4\textwidth]{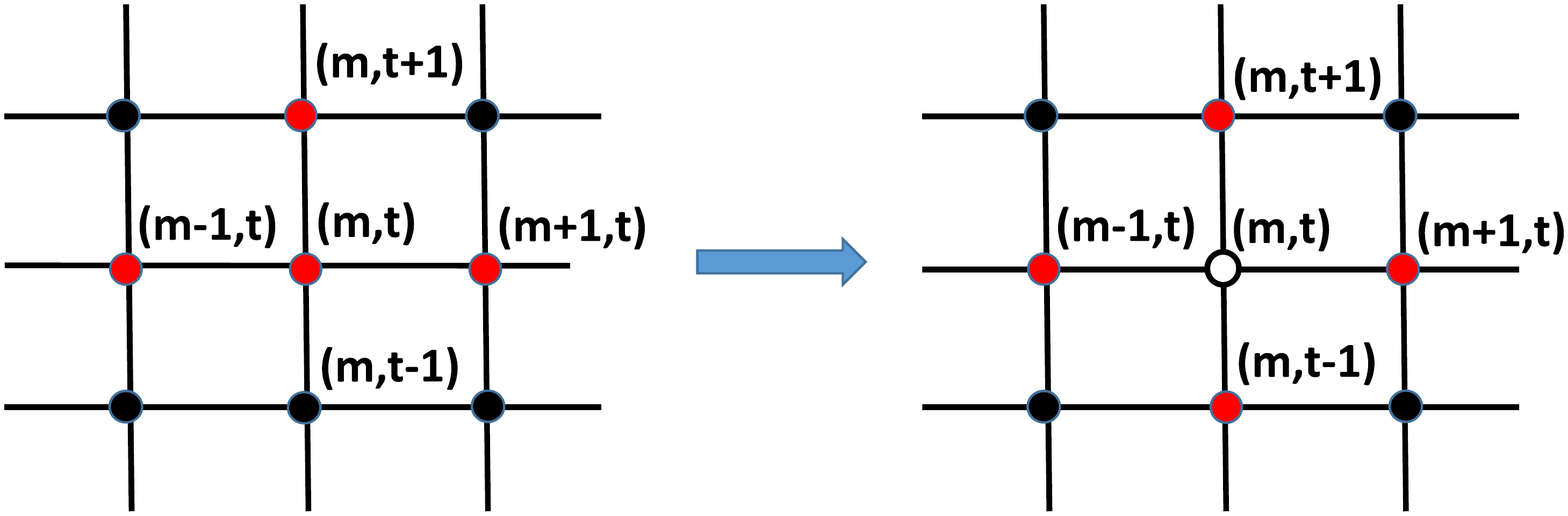}
   \vskip -2cm
   \includegraphics[width=0.4\textwidth]{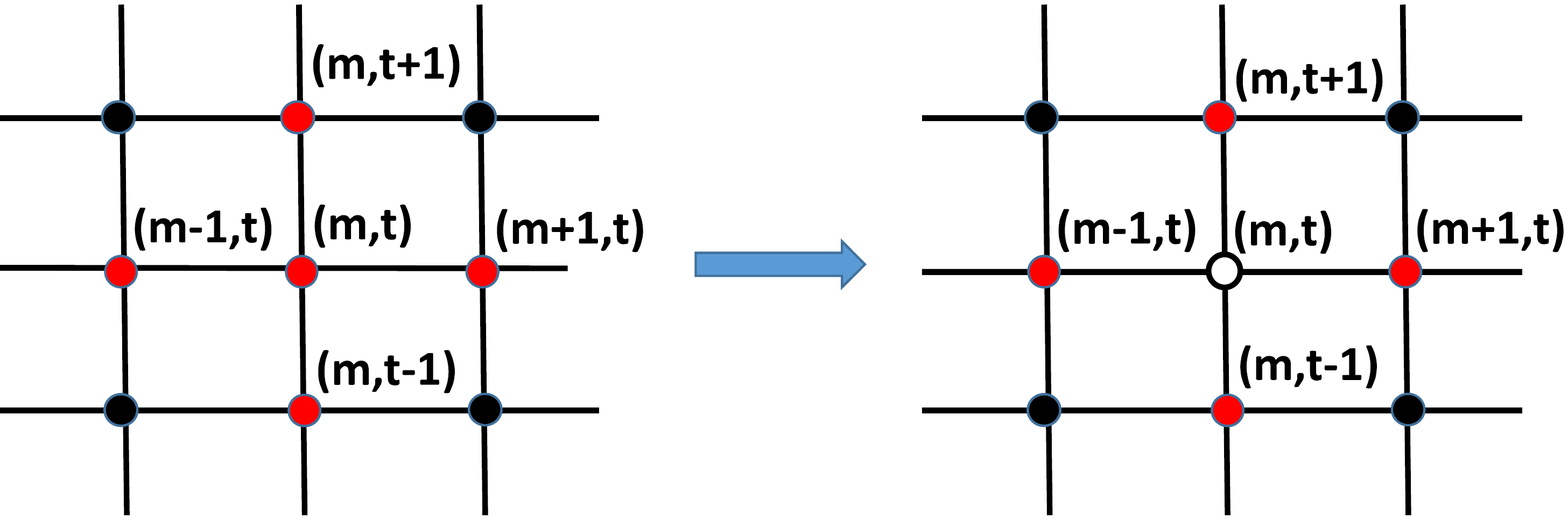}
    \caption[ Ku] {The figure illustrates contraction rules for lattice sites $(m,t)$ belonging to the set  $\mathcal{L}_2$. The four figures above correspond to the case where three out of four neighbours of $(m,t)$ belong to the set $\mathcal{L}_2$. The bottom figure  illustrates the case where all four neighbours  belong to $\mathcal{L}_2$. }
    \label{fig:DualMethod}
\end{figure}

Correlation function  between a number of  local  observables in the   Flouquet  KC (\ref{KickedChain})   can be written in the form of partition function, 
\begin{equation} 
Z=\frac{1}{L^{Nt}}
    \sum_{\{s_{m,k}| (m,k)\in\mathcal{L}_1\}}
   \left[ e^{-i\mathcal{F}(\{s_{m,k}\})} \prod_{(m,k)\in\mathcal{L}_2} \delta(s_{m,k},s_{m,1-k+2t})\right]  D(s_{z_1},\dots  ,s_{z_n}),\label{method1}
\end{equation}
where the last factor, $D$  depends on    a finite number of lattice sites, $\mathcal{L}_0=\{z_1,\dots,  z_l\}$   corresponding to location of the  observables.  The above sum, in general, runs over a subset   $\mathcal{L}_1$ of sites from  the $2t\times N$ lattice $\LNT=\{(m,k)| k=1, \dots ,2t, m=1,\dots,N\}$ while  the product in (\ref{method1}) is, furthermore,  restricted to a  subset   $\mathcal{L}_2 \subseteq \mathcal{L}_1$.  In what follows we will distinguish between  three type of points    $(m,k)\notin\mathcal{L}_0$   of  the  spatial-temporal lattice $\LNT$ and introduce the corresponding  symbolic  notation for lattice sites:  

\begin{itemize}
    \item \textit{Type 1:} $(m,k)\notin\mathcal{L}_1 $ i.e.,  there is no summation over the variables $s_{m,k},s_{m, 1-k+2t}$ in the partition function. The sites of this type are depicted by empty circles $\{\Circle\}$.
    \item \textit{Type 2:} $(m,t)\in\mathcal{L}_2 $  i.e., there is  summation over the variables $s_{m,k},s_{m,1-k+2t}$ coupled by the term $\delta(s_{m,k},s_{m,1-k+2t })$. The sites of this type are depicted by full red  circles $\{{\color{red}\CIRCLE}\}$.

    \item \textit{Type 3:} $(m,k)\in\mathcal{L}_1 \backslash \mathcal{L}_2 $ i.e.,  there is  summation over  uncoupled variables $s_{m,k},s_{m,1-k+2t}$. The sites of this type are depicted by full black  circles $\{\CIRCLE\}$.
\end{itemize}
Having this notation at hand, we can uniquely encode a partition function of the type (\ref{method1}) by filling nodes $(m,k)$  of the lattice $\mathcal{L}_1 \backslash \mathcal{L}_0 $   with  symbols drawn  from the alphabet  $\{\Circle, {\color{red}\CIRCLE},\CIRCLE\}$, see   figs.~\ref{fig:four_point1},\ref{fig:four_point2}. 

Thanks to   the unitarity  of the operator  $u_2$      a  simple graphical  method  for  calculation  of partition functions  like  (\ref{method1}) can be developed. To this end we establish ``contraction  rules"  for sites of $\LNT\backslash \mathcal{L}_0 $. 
Let $(m,k)$ be a site  of the type II  such that  three  of its  neighbours  are of the type II, and the forth one  of the type III. It can be easily shown that after summation over  $s_{m,k},s_{m,1-k+2t}$ variables the  fourth site becomes of the type II as well, while $ (m,k)$ becomes of the type I, see fig.~\ref{fig:DualMethod}.  Indeed, whenever   $(m-1,k), (m+1,k), (m,k), (m,k-1)\in \mathcal{L}_2$ we have for sum over  $s_{m,k},s_{m,1-k+2t}$ variables  in eq.~(\ref{method1})
\begin{equation}
    \frac{1}{L}\sum_{s_{m,k}} \sum_{s_{m,1-k+2t}} e^{-i(f_2(s_{m,k},s_{m,k+1})-f_2(s_{m,1-k+2t},s_{m,-k+2t}))}\delta(s_{m,k},s_{m,1-k+2t}) = \delta(s_{m,k+1},s_{m,-k+2t}).
\end{equation}
In an analogous way one can obtain all other contraction rules illustrated on fig.~\ref{fig:DualMethod}.  
Note that the above contraction rules    are   akin of the operator ``fusion rules"  introduced in \cite{BeKoPr19-4}.

Obviously, each contraction leads to removing of  two summation variables from the sum (\ref{method1}) without changing its form.   As a result, by consecutive  applications    of the contraction rules  the initial partition function  can be reduced to   the state where  the vast majority of the summation variables are excluded from the sum (\ref{method1}). The remaining sum can be then represented with the help of a transfer operator of  a small dimension, independent of $N$.

\subsection{Correlations between  operators with two-point support}

Here we consider the two point correlator,
\begin{equation}
    C(n,t)=L^{-N}\Trace U^{t} \Sigma_{n_1} U^{-t} \Sigma_{n_2 }, \qquad n=n_2-n_1,\label{TwoLegCorr}
\end{equation}
where operators $\Sigma_{n_1},\Sigma_{n_2}$ are given by 
eqs.~(\ref{operator1},\ref{operator2}). As the first step, we  cast  (\ref{TwoLegCorr}) into  the form of  partition function for a classical statistical model. Specifically,  we have 
\begin{multline}
    C(n,t)
    =\frac{1}{L^{Nt}}
    \sum_{\{s_{m,k}\in 1,\dots, L\}}
    e^{-i\mathcal{F}(\{s_{m,k}\})}
     \langle s_{n_1,2t}| \q_1^c |s_{n_1,1}\rangle \langle s_{n_1,2t}| \q_2^c |s_{n_1,1}\rangle
\langle s_{n_2,t}| \q_3 |s_{n_2,t+1}\rangle\langle s_{n_2,t}| \q_4 |s_{n_2,t+1}\rangle\\
    \times\prod_{m\neq n_1,n_1+1}^{N}\delta(s_{m,2t},s_{m,1})\prod_{m\neq n_2,n_2+1}^{N}\delta(s_{m,t},s_{m,t+1}), \label{part_func1}
\end{multline}
where
\begin{multline}
\mathcal{F}=\sum_{k=1}^t \sum_{m=1}^{N} f_1(s_{m,k}, s_{m+1,k}) - f_1(s_{m,k+t}, s_{m+1,k+t})
    +f_2(s_{m+1,k}, s_{m,k}) - f_2(s_{m,k+t}, s_{m+1,k+t}).\label{part_func2}
\end{multline}

We will consider  (\ref{TwoLegCorr})  at the cone light border  $n=t$.
It is instructive to represent  $C_t=C(n=t,t)$  in the form of a partition function. The initial expression   is shown in  a graphic form  on  fig.~ \ref{fig:four_point1}. The summation variables $s_{m,k}$ are excluded one by one by applying  the contraction rules, see  fig.~\ref{fig:DualMethod}. For $N>2t$ the  elimination of $s_{m,k}$ variables can be continued    up to reaching the stage illustrated by the  figure (\ref{fig:four_point2}).  Here the remaining  summation variables (shown in red and black) are located along the one dimensional strip only, which reduces the whole  problem to calculation of a quasi one-dimensional partition function. By using the transfer operator (\ref{transferOp}),  it can be cast into the form of the expectation value
\begin{equation}
    C_t=\langle\bar{\Phi}_{\q_1 \q_2}|\T^{t-2}|\Phi_{\q_3 \q_4}\rangle,\label{four_point_corr}
\end{equation}
where the left $\bar{\Phi}_{\q_1\q_2}$ and  the right  $\Phi_{\q_3\q_4}$  vectors are defined by  (\ref{vectors1}, \ref{vectors2}),    respectively.

\subsection{Application to KIC model}

The KIC model provides  a minimal realisation of the model  (\ref{KickedChain})   with $L=2$. The  KIC evolution is governed by the  Hamiltonians:
\begin{equation}
 \HI=\sum_{n=1}^N J\hat\sigma_n^z \hat\sigma_{n+1}^z +h\hat\sigma_n^z,\qquad 
\HK=b\sum_{n=1}^N  \hat\sigma_n^x,   \label{KICHamiltonian3} 
\end{equation}
\[ \hat\sigma_n^\alpha=\underbrace{\mathds{1}\otimes  \dots\otimes \mathds{1}}_{n -1} \otimes\,  \sigma^\alpha\otimes \underbrace{\mathds{1}\otimes  \dots\otimes \mathds{1}}_{N-n },\]
where $\sigma_n^\alpha, \alpha=x,y,z$ are Pauli matrices.  
 For the sake of simplicity of exposition we restrict our considerations to  $b=\pi/4$ and arbitrary $J,h$. Note that the dual-unitary case corresponds to $J=b=\pi/4$.  
The  resulting  evolutions  $\UK$, $\UI$  take the form  (\ref{eq:basePropagator}) with the functions  
\[f_1=- Jmn - \frac{h}{2}(m+n),  \qquad
f_2=\frac{\pi}{4}(mn-1) , \]
$m,n=\pm 1$, defining the two unitary matrices $u_1,u_2$:
\begin{equation}
    u_1=\frac{1}{\sqrt{2}}\begin{pmatrix} 
  e^{-i(J+h)} &   e^{iJ}   \\
  e^{iJ} &   e^{-i(J-h)} 
\end{pmatrix}, \qquad  u_2=\frac{1}{\sqrt{2}}\begin{pmatrix} 
  1 &  -i  \\
-i &  1 
\end{pmatrix}. \label{u1u2KIC}
\end{equation}

After inserting $f_1, f_2$ into eq.~(\ref{transferOp}) we obtain:

\begin{equation}
    \T=\frac{1}{2}  \left( \begin{array}{llll}
  \cos^2 h_+ &  \sin^2 h  & \sin^2 h_+  & \cos^2 h\\
  \sin^2 h_+  & \cos^2 h & \cos^2 h_+ &  \sin^2 h \\
    \sin^2 h  & \cos^2 h_- & \cos^2 h &  \sin^2 h_- \\
\cos^2 h &  \sin^2 h_-  & \sin^2 h  & \cos^2 h_-
\end{array}\right), \label{transferKIC1}
\end{equation}
where $ h_+=h+J-\pi/4$, $ h_-=h-J+\pi/4$. The four eigenvalues of  $\T$  are \[\mu_1=1, \qquad \mu_2=\cos 2h \sin^2 2J,\qquad \mu_3=0, \qquad \mu_4=0.\]
As a result, the  $n$-th power of $\T$ is given for $n>1$ by 
\begin{equation}
\T^n=\mu_2^n \, \Phi_2\otimes \bar\Phi_2  +     \Phi_1\otimes \Phi_1 \label{powerofT}
\end{equation}
with $\Phi_1=\frac{1}{2}(1,1,1, 1)^T$ being the eigenvector of ${\T}$ for the leading  eigenvalue  $\mu_1$ and 
\begin{equation}
   \Phi_2=\frac{1}{c+d}(c,-c,-d, d)^T \qquad \bar\Phi_2=\frac{1}{c+d}(c,-d,-c, d),  \label{vectorsPhi2}
\end{equation}
$c=\cos 2h_+ +\cos 2h,  d=\cos 2h_- +\cos 2h$, are the left and right eigenvectors corresponding to $\mu_2$.

\subsubsection{ Four-point correlators.}

To evaluate correlators note that the  operators  $u_2 {\q}_1 u^\dagger_2, \q_4 $  contribute only diagonal elements into  (\ref{vectors1},\ref{vectors2}). In the case  of  KIC model this means that only the spin  combinations,   $\Sigma_{n_1}=\hat\sigma^{\alpha}_{n_1} \hat\sigma^\beta_{n_1+1}$,   $\Sigma_{n_2}=\hat\sigma^\gamma_{n_2} \hat\sigma^{\delta}_{n_2+1}$ for $\alpha=y, \delta=z$ might have  $C_t\neq 0$.
By using the representation (\ref{powerofT}) we have for the correlator (\ref{four_point_corr5})
\begin{equation}
 C_t\equiv C(t,t)=\mu_2^{t-2}\langle\bar{\Phi}_{\sigma^y\sigma^\beta}| \Phi_2\rangle\langle\bar\Phi_2|\Phi_{\sigma^\gamma\sigma^z}\rangle+\langle\bar{\Phi}_{\sigma^y\sigma^\beta}| \Phi_1\rangle\langle\Phi_1|\Phi_{\sigma^\gamma\sigma^z}\rangle,  \label{Corr11}
\end{equation}
where the vectors $\bar{\Phi}_{\sigma^y\sigma^\beta}, \Phi_{\sigma^\gamma\sigma^z}$ are  calculated by  (\ref{vectors1}, \ref{vectors2}). Explicitly, they  are given by 
\begin{equation}
  \Phi_{\sigma^y \sigma^z}=\bar{\Phi}_{\sigma^y \sigma^z}=\frac{\sin 2J}{2} \left( \begin{array}{r} 
  -\sin (2h-2J)\\
   \sin(2h-2J) \\
     -\sin (2h+2J) \\
\sin(2h+2J)
\end{array} \right), \quad
 \Phi_{\sigma^x\sigma^z}= \bar{\Phi}_{\sigma^y\sigma^x}=\frac{\sin 2J}{2} \left( \begin{array}{r}
  -\cos (2h-2J)\\
   -\cos(2h+2J) \\
     \cos (2h-2J) \\
\cos(2h+2J)
\end{array} \right).  \label{vectorsPhi4}
\end{equation}
After inserting  (\ref{vectorsPhi2},\ref{vectorsPhi4}) into  (\ref{Corr11}) we  obtain 
\begin{equation}
C_t=\mathcal{C}_{\alpha\beta}^{\gamma\delta}(\cos  2h\sin^2 2J)^t,\label{main_result43}
 \end{equation}
 where prefactors,  $\mathcal{C}_{\alpha\beta}^{\gamma\delta}$ are given by
 \begin{equation}
\mathcal{C}_{yz}^{yz}=1,\qquad  \mathcal{C}_{yx}^{xz}=\tan^2 2h, \qquad
 \mathcal{C}_{yz}^{xz}=\mathcal{C}_{yx}^{yz}=-\tan 2h
 \end{equation} while   zeroes for all other spin combinations.

\subsubsection{ Two-point correlators.}

By using the representation (\ref{powerofT}) we have for the correlator (\ref{four_point_corr2})
\begin{equation}
 C^{\alpha \beta}(t-1,t)=\mu_2^{t-2}\langle\bar{\Phi}_{\mathds{1}\sigma^\alpha}| \Phi_2\rangle\langle\bar\Phi_2|\Phi_{\sigma^\beta\mathds{1}}\rangle+\langle\bar{\Phi}_{\mathds{1}\sigma^\alpha}| \Phi_1\rangle\langle\Phi_1|\Phi_{\sigma^\beta\mathds{1}}\rangle,  \label{Corr12}
\end{equation}
where the vectors $\bar{\Phi}_{\mathds{1}\sigma^\alpha}, \Phi_{\sigma^\beta\mathds{1}}$ can be calculated by  (\ref{vectors1}, \ref{vectors2}). Explicitly, they  are given by 
\begin{equation}
  \Phi_{\sigma^y\mathds{1}}=\frac{\cos 2J}{2} \left( \begin{array}{r} 
  \cos (2h-2J)\\
   -\cos(2h-2J) \\
     \cos (2h+2J) \\
-\cos(2h+2J)
\end{array} \right), \quad
 \Phi_{\sigma^x\mathds{1}}=\frac{\cos 2J}{2} \left( \begin{array}{r}
  -\sin (2h-2J)\\
   -\sin(2h+2J) \\
     \sin (2h-2J) \\
\sin(2h+2J)
\end{array} \right) ,
\quad \Phi_{\sigma^z\mathds{1}}=0, \label{vectorsPhi3}
\end{equation}
and $\Phi_{\sigma^y\mathds{1}}= \bar\Phi_{\mathds{1}\sigma^z}$, $\Phi_{\sigma^z\mathds{1}}= \bar\Phi_{\mathds{1}\sigma^y}$,   $\Phi_{\sigma^x\mathds{1}}= \bar\Phi_{\mathds{1}\sigma^x}$. After substitution of  (\ref{vectorsPhi3}) into eq.~(\ref{Corr12})  one has 
\begin{equation}
 C^{\alpha \beta}(t-1,t)= \mathcal{C}^{\alpha \beta} (\cos 2h \sin^{2} 2J)^t\cot^2 2J \label{KICCorrelator2}
 \end{equation}
 with the coefficients given by 
 \[
\mathcal{C}^{xx} = 1,\qquad \mathcal{C}^{xy} = \mathcal{C}^{zx} = \tan 2h,\qquad \mathcal{C}^{zy} = \tan^2 2h,
\]
and  zeroes for all  other $\alpha,\beta$  combinations. 
\end{document}